\documentclass[aip,jcp,reprint,eqsecnum,superscriptaddress,twocolumn,longbibliography,floatfix]{revtex4-2}

\usepackage{empheq}
\usepackage{graphicx}
\usepackage{dcolumn}
\usepackage{bm}

\usepackage[utf8]{inputenc}
\usepackage[T1]{fontenc}
\usepackage{mathptmx}
\usepackage{etoolbox}
\usepackage{xcolor}
\usepackage{mathpazo}
\usepackage{bm}
\usepackage[unicode=true,
 bookmarks=true,bookmarksnumbered=false,bookmarksopen=false,
 breaklinks=false,pdfborder={0 0 0},pdfborderstyle={},backref=false,colorlinks=true]
 {hyperref}
\hypersetup{pdfborderstyle={},pdfborderstyle={},pdfborderstyle={},pdfborderstyle={},pdfborderstyle={},pdfborderstyle={},linkcolor=blue,citecolor=blue}
\usepackage{breakurl}

\usepackage{natbib}
\usepackage{amssymb,amsmath}

\newcommand{\cl}{\text{cp}}
\newcommand{\MM}{\text{irr}}
\newcommand\beq{\begin{equation}}
\newcommand\eeq{\end{equation}}
\newcommand\beqa{\begin{eqnarray}}
\newcommand\eeqa{\end{eqnarray}}
\newcommand{\nn}{\nonumber\\}
\def\bal#1\eal{\begin{align}#1\end{align}}

\newcommand{\xx}{5}
\newcommand{\yy}{5}
\newcommand{\xmax}{30}
\newcommand{\ymax}{22}

\newcommand{\Stwo}{\begin{picture}(\xmax,\ymax)(-\xx,\yy)
\setlength{\unitlength}{.1mm}
\put(0,30){\circle*{18}}
\put(60,30){\circle*{18}}
\put(9,30){\line(1,0){42}}
\end{picture}}

\newcommand{\Rthree}{\begin{picture}(\xmax,\ymax)(-\xx,\yy)
\setlength{\unitlength}{.1mm}
\put(0,0){\circle*{18}}
\put(60,0){\circle*{18}}
\put(30,60){\circle*{18}}
\put(4,8){\line(1,2){22}}
\put(56,8){\line(-1,2){22}}
\end{picture}}

\newcommand{\Sthree}{\begin{picture}(\xmax,\ymax)(-\xx,\yy)
\setlength{\unitlength}{.1mm}
\put(0,0){\circle*{18}}
\put(60,0){\circle*{18}}
\put(30,60){\circle*{18}}
\put(4,8){\line(1,2){22}}
\put(9,0){\line(1,0){42}}
\put(56,8){\line(-1,2){22}}
\end{picture}}

\makeatletter
\def\@email#1#2{%
 \endgroup
 \patchcmd{\titleblock@produce}
  {\frontmatter@RRAPformat}
  {\frontmatter@RRAPformat{\produce@RRAP{*#1\href{mailto:#2}{#2}}}\frontmatter@RRAPformat}
  {}{}
}%
\makeatother
\begin{document}

\title[Hard-disk fluids under single-file confinement]{Equation of state of hard-disk fluids under single-file confinement}
\author{Ana M. Montero}
\affiliation{Departamento de F\'isica, Universidad de Extremadura, E-06006 Badajoz, Spain}
\email{anamontero@unex.es}
\author{Andr\'es Santos}%
 \affiliation{
Departamento de F\'isica, Universidad de Extremadura, E-06006 Badajoz, Spain
and Instituto de Computaci\'on Cient\'ifica Avanzada (ICCAEx), Universidad de Extremadura, E-06006 Badajoz, Spain
}%
\email{andres@unex.es}

\date{\today}

\begin{abstract}
The exact transfer-matrix solution for the longitudinal equilibrium properties of the single-file hard-disk fluid is used to study the limiting low- and high-pressure behaviors analytically as functions of the pore width. In the low-pressure regime, the exact third and fourth virial coefficients are obtained, which  involve single and double integrals, respectively. Moreover, we show that the standard irreducible diagrams do not provide a complete account of the virial coefficients in confined geometries. The asymptotic equation of state in the high-pressure limit is seen to present a simple pole at the close-packing linear density, as in the hard-rod fluid, but, in contrast to the latter case, the residue is $2$. Since, for an arbitrary pressure, the exact  transfer-matrix treatment requires the numerical solution of an eigenvalue integral equation, we propose here two simple approximations to the equation of state, with different complexity levels, and carry out an extensive assessment of their validity and practical convenience vs the exact solution and available computer simulations.
\end{abstract}

\maketitle

\section{\label{sec:introduction}Introduction}
Confined fluid systems are an important field of study due to the great range of applications and situations where they can be found. Physically interesting systems in biology or chemistry involve dealing with confined particles, such as carbon nanotubes\cite{Ketal11,MCH11} or biological ion channels,\cite{DNHEG08} to cite just a couple of examples. In many of these systems, the geometry is so restrictive that they become quasi-one-dimensional (Q1D) systems.

These Q1D systems can  be used to model a wide range of extremely confined two- or three-dimensional systems, in which the space available along one of the dimensions is much larger than that along the other ones. The study of this type of fluids is especially interesting from a statistical--mechanical perspective since many of them are amenable to exact analytical solutions, therefore providing insight into the thermodynamic and structural properties of such systems. An important subset of confined fluids is  made of those under the so-called single-file confinement,\cite{PGIB21,HP18} where particles are inside a pore that is not wide enough to allow particles to either bypass each other or interact with their second nearest neighbors, therefore confining them into a single-file formation.

Q1D systems, usually restricted to single-file configurations,  constitute an active field of study for both equilibrium\cite{B62,B64b,WPM82,PK92,KP93,P02,KMP04,FMP04,VBG11,GV13,GM14,GM15,HFC18,M14b,M15,M20,HBPT20,P20,PBT22,JF22}  and nonequilibrium properties,\cite{GM14,FMP04,KMS14,RGM16,TFCM17,WLB20,LG20,HBPT21,RS23,MBGM22,RGIB23,MGB22} as well as for jamming effects,\cite{GM14,ZGM20,I20,LM20,LM22} from different perspectives.
In the case of confined two-dimensional (2D) systems, a simple but, nevertheless, functional way of modeling the particle interaction is by means of the hard-disk interaction potential, in which particles are not allowed to interpenetrate but otherwise they do not interact among themselves.

It is important to bear in mind that only the most relevant (longitudinal) thermodynamic properties of the original confined 2D fluid are mapped onto those of the effective Q1D system. In this sense, Barker's solution\cite{B62,B64b} for the single-file configuration with only nearest-neighbor interactions was based on an averaged potential function for the disk--disk interactions. A perhaps more insightful solution was found by Kofke and Post via the transfer-matrix method.\cite{KP93} Most of the subsequent theoretical studies\cite{P02,GM14,HFC18,M14b,M15,M20} also focused on the physical properties of the effective Q1D system, while in other works, the transverse properties of the genuine 2D fluid were analyzed as well.\cite{FMP04,VBG11,GV13,GM15,HBPT20,P20,PBT22,JF22} In particular, an exact analytical canonical partition function for the 2D system has recently been obtained.\cite{P20}
Even if the theoretical advances refer to the effective Q1D system, their validity needs to be tested against computer simulations on the original 2D system.\cite{KP93,VBG11,GM14,M20,HBPT20,JF22}

The exact transfer-matrix thermodynamic solution for the Q1D fluid\cite{KP93} involves numerical schemes to solve an eigenvalue equation in order to obtain the equation of state of the system, and no analytical solution has yet been found. In this sense, several proposals have been developed during the last few years to obtain analytically accurate approximations to the exact solution, involving first-order approximations of the contact distance of the particles,\cite{VBG11} virial-coefficient expansions,\cite{M14b,M15,M20} or distinguishing between high- and low-pressure regimes.\cite{KMP04,GM14}

In this paper, we revisit the exact transfer-matrix solution\cite{KP93} for the single-file Q1D hard-disk fluid and perform a perturbation analysis to calculate the exact third and fourth virial coefficients.
Interestingly, they differ  from previous evaluations via the standard diagrammatic method,\cite{M14b,M15,M20} the reason being that the textbook cancellation of the so-called \emph{reducible} diagrams does not hold in the case of confined fluids. We also study the behavior in the  high-pressure limit,  finding that the residue of the simple pole at close packing differs from that in the pure (1D) hard-rod system.  In view of this, we propose two different analytical approximations for the equation of state and study their behavior against the exact solution and available computer simulations. Despite its simplicity, our basic uniform-profile approximation   recovers the second virial coefficient, provides reasonable estimates of the third and fourth virial coefficients, and predicts the correct close-packing linear density. A more sophisticated (and accurate) exponential-profile approximation improves the estimates of the third and fourth virial coefficients,  reduces to the exact solution in the close-packing limit, and exhibits an excellent behavior for intermediate densities. Moreover, the execution times of the uniform-profile and exponential-profile approximations are seen to be up to about $10^5$ and $10^3$ times shorter, respectively, than the exact solution for high pressures and wide pores.

Our paper is organized as follows: Sec.~\ref{sec:solution} defines the system and its exact solution, including an analysis of the low- and high-pressure behaviors in Secs.~\ref{sec:system_lowpressure} and~\ref{sec:system_highpressure}, respectively. Section~\ref{sec:approximations} presents our two analytical approximations to the equation of state, while  an assessment of both approximations vs the exact solution is carried out in Sec.~\ref{sec:4}. This paper is closed in Sec.~\ref{sec: concl} with some concluding remarks. The most technical details are relegated to Appendices~\ref{app:mapping_onedimension}--\ref{app:numerical_methods}.

\section{\label{sec:solution}The Confined Hard-Disk Fluid. Exact Properties}

\subsection{System}

\begin{figure}
    \includegraphics[trim={0cm 3cm 0cm 3cm},clip,width=0.95\columnwidth]{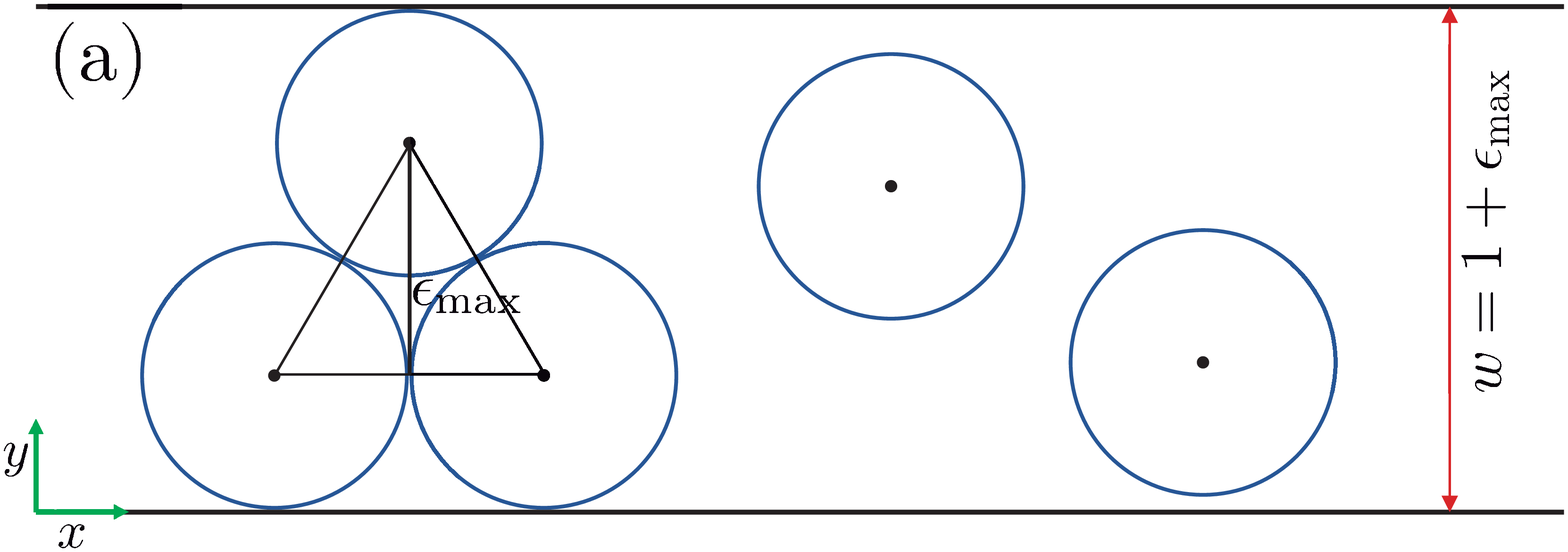}\\
    \includegraphics[trim={0cm 3cm 0cm 3cm},clip,width=0.95\columnwidth]{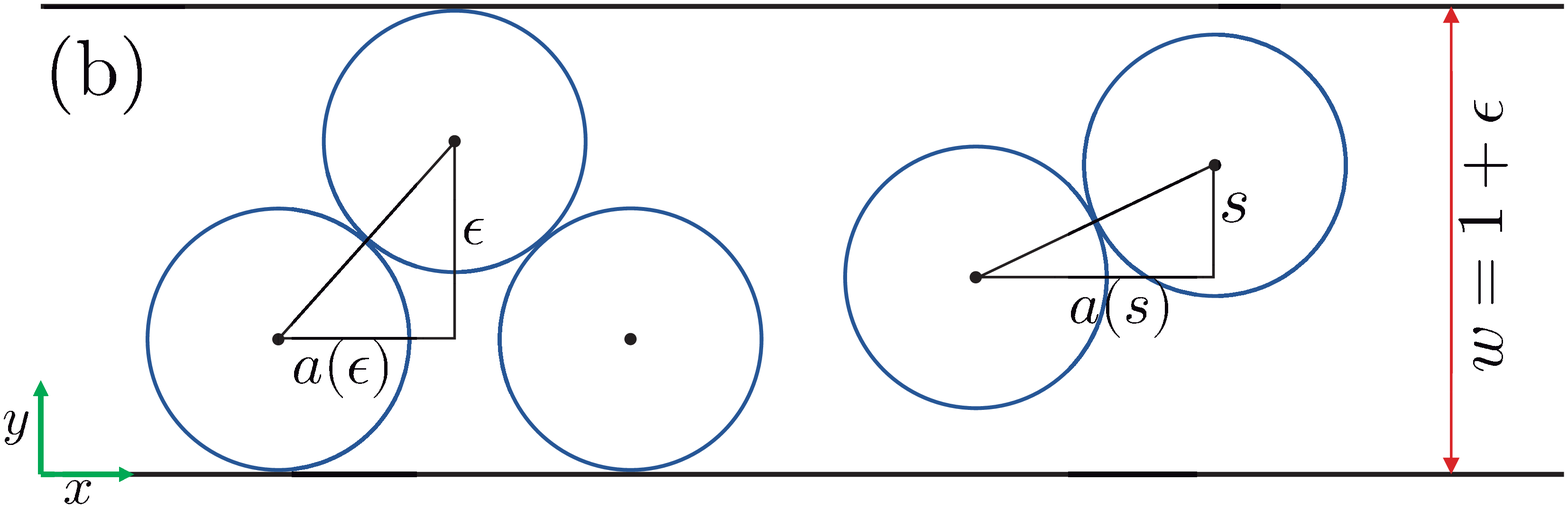}
    \caption{Schematic representation of the single-file hard-disk fluid. Panel (a) shows the maximum allowed value of the pore size, $1+\epsilon_{\max}$ (with $\epsilon_{\mathrm{max}}=\sqrt{3}/2$),  beyond which a disk can interact with its second nearest-neighbors, thus violating the single-file condition. Panel (b) depicts a case with $\epsilon<\epsilon_{\mathrm{max}}$, where the two disks on the right show the definition of the longitudinal separation at contact, $a(s)$, while the three disks on the left  illustrate the close-packing configuration.}
    \label{fig:model_image02}
\end{figure}

We consider a system of $N$  hard disks of unit diameter confined in a long channel of length $L \gg 1$ and width $w = 1+\epsilon$, with $0\leq \epsilon \leq \epsilon_{\mathrm{max}}$, where
$\epsilon_{\mathrm{max}}=\sqrt{3}/2\simeq 0.866$ in order to ensure the single-file condition and preclude second nearest-neighbor interactions, as depicted in Fig.~\ref{fig:model_image02}(a).
As illustrated in Fig.~\ref{fig:model_image02}(b), if the transverse separation between two disks at contact is $s$, their longitudinal separation is
\beq
\label{eq:a(s)}
a(s)\equiv\sqrt{1-s^2}.
\eeq

The number of disks per unit area is $\rho=N/Lw$. However, in the  Q1D configuration of the system, it is convenient to characterize the number density through the number of particles per unit length, $\lambda \equiv N/L=\rho w$. Its close-packing value (given an excess pore  width $\epsilon$) is $\lambda_{\mathrm{cp}}(\epsilon) = 1/a(\epsilon)$, as inferred from Fig.~\ref{fig:model_image02}(b),  at which the particles occupy the maximum available space, resulting in the pressure diverging at that value. This divergence will be discussed in depth in Sec.~\ref{sec:system_highpressure}. We note that $\lambda_{\text{cp}}(\epsilon_{\mathrm{max}}) = 2$.

Due to the anisotropy of the original 2D system, the transverse pressure ($P_\perp$) is different from the longitudinal one ($P_{\|})$. We, then,  define the (reduced) Q1D pressure as $p \equiv P_{\|} w$, where, henceforth, we take $k_BT=1$ as unit energy ($k_B$ and $T$ being the Boltzmann constant and the absolute temperature, respectively).

\subsection{\label{sec:system_solution}Transfer-matrix solution}
The exact solution to the Q1D system can be obtained via the transfer-matrix method. In the thermodynamic limit of large $N$, the excess Gibbs free energy per particle, $g^{\text{ex}}(p)$, may be written as\cite{KP93}
\begin{equation}\label{eq:gibbs_equation}
     g^{\text{ex}}(p) =  - \ln\frac{\ell(p)}{\epsilon},
\end{equation}
where $\ell(p)$ is the maximum eigenvalue corresponding to the problem
\begin{equation}\label{eq:eigenvalue_equation}
    \int\mathrm{d}y_2\, e^{-a(y_1-y_2)p}\phi(y_2)=\ell \phi(y_1),
\end{equation}
and $\phi(y)$ is the associated eigenfunction. Here and henceforth, all integrations over the $y$-variable will be understood to run along the interval $-\epsilon/2\leq y\leq \epsilon/2$ (where the origin $y=0$ is taken at the centerline) and the integration limits will be omitted.
Under the normalization condition
\begin{equation}\label{eq:eigenvalue_normalization}
    \int\mathrm{d}y\,\phi^2(y)=1,
\end{equation}
$\phi^2(y)$ represents  the probability density along the transverse direction $y$ within this framework.
Multiplying both sides of Eq.~\eqref{eq:eigenvalue_equation} by $\phi(y_1)$ and integrating over $y_1$, we obtain
\begin{equation}
\label{eq:2.5}
    \ell = \int\mathrm{d}y_1\int\mathrm{d}y_2\, e^{-a(y_1-y_2)p}\phi(y_1)\phi(y_2),
\end{equation}
where the normalization condition, Eq.~\eqref{eq:eigenvalue_normalization}, has been used.

Of course, both $\ell$ and $\phi(y)$ are functions of $p$. Differentiating both sides of Eq.~\eqref{eq:2.5} with respect to $p$, one obtains
\bal
\label{eq:ell_derivative}
\partial_p\ell=&-\int\mathrm{d}y_1\int\mathrm{d}y_2\, e^{-a(y_1-y_2)p}a(y_1-y_2)\phi(y_1)\phi(y_2)\nn
&+2\int\mathrm{d}y_1\int\mathrm{d}y_2\, e^{-a(y_1-y_2)p}\phi(y_2)\partial_p\phi(y_1).
\eal
On account of Eq.~\eqref{eq:eigenvalue_equation}, the second term on the right-hand side of Eq.~\eqref{eq:ell_derivative} can be rewritten as $2\ell \int\mathrm{d}y_1\,\phi(y_1) \partial_p \phi(y_1)=\ell \partial_p\int\mathrm{d}y_1\,\phi^2(y_1)=0$. Thus, $\partial_p\ell$ is only given by the first term on the right-hand side of Eq.~\eqref{eq:ell_derivative}.

The compressibility factor $ Z \equiv p/\lambda$ can be obtained from the Gibbs free energy by the thermodynamic relation $Z=1+p\partial_p  g^{\text{ex}}(p)=1-(p/\ell)\partial_p\ell$. Making use of Eq.~\eqref{eq:ell_derivative}, one obtains
\begin{equation}\label{eq:z_exact_01}
    Z = 1 + \frac{p}{\ell} \,\int\mathrm{d}y_1\int\mathrm{d}y_2 \,e^{-a(y_1-y_2)p}a(y_1-y_2)\phi(y_1)\phi(y_2).
\end{equation}
Taking into account  Eq.~\eqref{eq:2.5}, Eq.~\eqref{eq:z_exact_01} can be rewritten as
\begin{equation}\label{eq:z_exact_02}
      Z = 1 + p \frac{\int\mathrm{d}y_1\int\mathrm{d}y_2 \, e^{-a(y_1-y_2)p}a(y_1-y_2)\phi(y_1)\phi(y_2)}{\int\mathrm{d}y_1\int\mathrm{d}y_2 \, e^{-a(y_1-y_2)p}\phi(y_1)\phi(y_2)}.
\end{equation}
It should be noted that, in contrast to the form \eqref{eq:z_exact_01}, the eigenfunction $\phi(y)$ in the form \eqref{eq:z_exact_02} does not need to be normalized. While both forms are fully equivalent inasmuch as the exact $\ell$ and $\phi(y)$ are used, they differ in the case of approximations.

It is interesting to remark that the solution shown here can also be obtained by a mapping of the original Q1D system onto a 1D non-additive mixture of hard rods, as outlined in Appendix~\ref{app:mapping_onedimension}.

It should be noted also that in the limit $\epsilon\to 0$ (at finite $p$),  one obtains  $\phi(y)\to \epsilon^{-1/2}\Theta(\frac{\epsilon}{2}-|y|)$, $\ell\to e^{-p}\epsilon$, $g^{\text{ex}}(p)\to p$, and $Z\to 1+p$ from Eqs.~\eqref{eq:eigenvalue_equation}, \eqref{eq:gibbs_equation}, and \eqref{eq:z_exact_01}, respectively, thus recovering the equation of state of the Tonks gas,\cite{T36} as expected.

\subsection{\label{sec:system_lowpressure}Low-pressure behavior}

Virial expansions are one of the most common methods to describe fluids under low-density (or, equivalently, low-pressure) conditions.\cite{HM13,S16}  In general, access to the exact virial coefficients of a given system, at least the lower-order ones, is fundamental to  improve the knowledge of the system and also to test the accuracy of approximate methods.

The virial coefficients $\{B_n\}$ are defined from the expansion of the compressibility factor in powers of density,
\beq
\label{eq:virial_density}
Z=1+\sum_{n=2}^\infty B_n \lambda^{n-1}.
\eeq
Analogously, one can introduce the expansion of $g^{\text{ex}}$ and $Z$ in powers of pressure,
\begin{subequations}
\beq
\label{eq:2.9a}
g^{\text{ex}}=\sum_{n=2}^\infty \frac{B_n'}{n-1} p^{n-1},
\eeq
\beq
\label{eq:2.9b}
Z=1+\sum_{n=2}^\infty B_n' p^{n-1},
\eeq
\end{subequations}
where
\begin{equation}
\label{eq:2.10}
    B'_2 = B_2,\quad B'_3 = B_3-B_2^2,\quad B'_4 = B_4-3 B_2 B_3 + 2B_2^3,
\end{equation}
and so on.
The second virial coefficient has an analytical expression, namely,\cite{KMP04,M18}
\begin{equation}
\label{eq:B2}
    B_2 = \frac{2}{3}\frac{ \left(1+\frac{\epsilon ^2}{2}\right) \sqrt{1-\epsilon ^2}-1}{\epsilon ^2}+\frac{\sin ^{-1}(\epsilon )}{\epsilon }.
\end{equation}

To the best of our knowledge, the correct third and fourth virial coefficients have not been evaluated yet. Here, we derive them from the exact transfer-matrix solution, Eq.~\eqref{eq:z_exact_01}, without assuming the direct application of the standard diagrammatic method.\cite{M14b,M15,M20}

Let us introduce the expansion in powers of $p$ of both the eigenvalue and the eigenfunction in Eq.~\eqref{eq:eigenvalue_equation} as
\begin{equation}\label{eq:virial_series}
\phi(y) = \sum_{n=0}^{\infty} \phi_n(y) p^n,
\quad
\ell = \sum_{n=0}^{\infty} \ell_n p^n.
\end{equation}
Inserting the expansion of $\ell$ into Eq.~\eqref{eq:gibbs_equation} and comparing with Eq.~\eqref{eq:2.9a}, we obtain
\beq
\label{eq:2.13}
B_3'=-2\frac{\ell_2}{\epsilon}+B_2^2,
\quad
B_4'=-3\frac{\ell_3}{\epsilon}-3B_2\frac{\ell_2}{\epsilon}+B_2^3,
\eeq
where we have used $\ell_0=\epsilon$ and $\ell_1=-\epsilon B_2$ (see Appendix~\ref{app:virial_series_math}).
Alternatively, the expansion of $\phi(y)$ provides the expansion of the integral
\bal
\label{eq:I}
I\equiv&\int\mathrm{d}y_1\int\mathrm{d}y_2\, e^{-a(y_1-y_2)p}a(y_1-y_2)\phi(y_1)\phi(y_2)\nn
=&\sum_{n=0}^\infty I_n p^n.
\eal
Since $I=-\partial_p\ell$ [see Eq.~\eqref{eq:ell_derivative}], one has
\beq
\label{eq:In}
I_n=-(n+1)\ell_{n+1}.
\eeq

By inserting the series expansions of Eq.~\eqref{eq:virial_series} into both the normalization condition, Eq.~\eqref{eq:eigenvalue_normalization}, and the eigenvalue equation, Eq.~\eqref{eq:eigenvalue_equation}, and equating the coefficients with the same powers of $p$ on both sides of the equation, one can, in principle, obtain as many terms  as desired. Appendix~\ref{app:virial_series_math} shows the calculation of  $\{\phi_0,\phi_1,\phi_2\}$ and $\{\ell_0,\ell_1,\ell_2\}$. In addition, $\ell_3$ can be obtained from $I_2$. Substitution of $\ell_2$ and $\ell_3$ into Eq.~\eqref{eq:2.13}, yields
\begin{subequations}
\label{eq:B3p&B4p}
\bal
\label{eq:B3p}
B_3'=&-\left(1+2W_2-3B_2^2-\frac{\epsilon^2}{6}\right)\nn
=&-\frac{\epsilon^4}{80}\left(1 +\frac{41 \epsilon^2}{126} + \frac{349 \epsilon^4}{2520}
+\cdots\right),
\eal
\bal
\label{eq:B4p}
B_4'=&-\Bigg[\left(12W_2-10B_2^2+\frac{3}{2}-\frac{\epsilon^2}{4}\right)B_2-3W_3\nn
&+\frac{(1-\epsilon^2)^{5/2}-1-5\epsilon^2}{15\epsilon^2}\Bigg]\nn
=&-\frac{23\epsilon^6}{15120}\left(1 +\frac{567 \epsilon^2}{920}+\frac{14823 \epsilon^4}{40480}
+\cdots\right),
\eal
\end{subequations}
where  $W_2$ and $W_3$ are given by  Eqs.~\eqref{eq:W_2} and \eqref{eq:W_3}, requiring to numerically carry out a simple and double integration, respectively.

The exact expressions derived here for $B_3'$ and $B_4'$ turn out to differ from those (hereafter referred to as $B_{3,\MM}'$ and $B_{4,\MM}'$) obtained via the integration of {standard}  \emph{irreducible} diagrams.\cite{M14b,M15,M20} In particular, the leading terms in the expansions in powers of $\epsilon$ of the latter coefficients are $B_{3,\MM}'=-\frac{\epsilon^4}{144}+\mathcal{O}(\epsilon^6)$ and $B_{4,\MM}'=-\frac{\epsilon^6}{160}+\mathcal{O}(\epsilon^8)$, which contrast with the leading terms in Eq.~\eqref{eq:B3p&B4p}.

The origin of the discrepancy between the exact virial coefficients obtained here from the transfer-matrix solution, Eq.~\eqref{eq:z_exact_01}, and those derived from the standard diagrammatic scheme\cite{M14b,M15,M20}  lies on the implicit assumption of a cancellation of the so-called \emph{reducible} diagrams in the latter method. This cancellation is inherently associated with the factorization property of the reducible diagrams into products of irreducible ones,\cite{S16} as a consequence of the translational invariance of the position of any particle. While this factorization property holds in bulk fluids, it fails under confinement, due to the breakdown of the translational invariance along the confined directions.

\begin{figure}
    \centering
    \includegraphics[width=0.78\columnwidth]{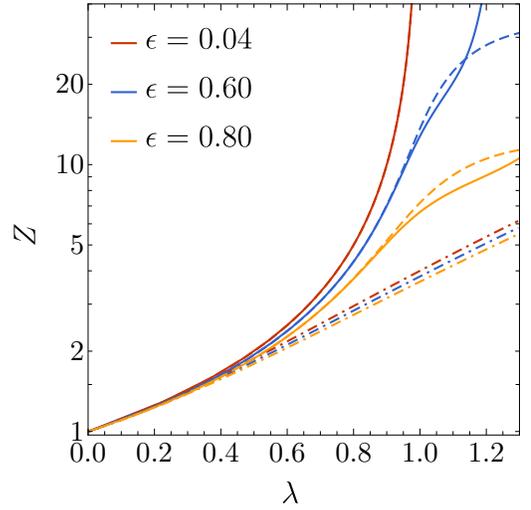}
    \caption{Comparison between the exact compressibility factor (solid lines), the truncated series $Z_{\text{tr}}(\lambda)=1+B_2\lambda+B_3\lambda^2+B_4\lambda^3$ (dashed-dotted lines), and the alternative truncated series $Z_{\text{tr}}'(p)=1+B_2 p+B_3'p^2+B_4' p^3$ (dashed lines) for the range $0\leq\lambda\leq 1.3$. The values of the pore width parameter are (from top to bottom) $\epsilon=0.04$, $0.6$, and $0.8$. On the scale of the figure, the results corresponding to $\epsilon=0.04$ are indistinguishable from those of the Tonks gas ($\epsilon=0$).\cite{T36}}
    \label{fig:z_truncated}
\end{figure}

Let us take the coefficient $B_3$ as the simplest example. By assuming cancellation of the reducible diagrams, one would have a single irreducible diagram, namely,\cite{M20}
\beq
B_{3,\MM}=-\frac{1}{3}\Sthree.
\eeq
On the other hand,  the actual result is
\beq
B_3=B_{3,\MM}+\Delta B_3,\quad \Delta B_3\equiv(\Stwo\!\!\!)^2-\Rthree.
\eeq
Here, the diagrams have its usual meaning,\cite{S16} except that they are supposed to be divided by $L\epsilon^n$, $n$ being the number of particles represented in the diagram.
In a bulk fluid, $\Delta B_3=0$, due to the factorization property of the reducible diagrams mentioned before. However, in our confined system, one has
\beq
\Stwo=-2B_2,\quad \Rthree=4W_2,
\eeq
so that $\Delta B_3=4(B_2^2-W_2)\neq 0$. As a by-product,  from Eq.~\eqref{eq:B3p}, we obtain
\beq
\label{eq:B3Mp}
B_{3,\MM}'=B_3'-\Delta B_3
=-\left(1-2W_2+B_2^2-\frac{\epsilon^2}{6}\right).
\eeq
This is equivalent to but much more compact than  the expression  derived in Ref.~\onlinecite{M20}.

It is worth mentioning that this issue regarding the correction needed to the irreducible-diagram representation of the virial coefficients arises also when dealing with flexible molecules.\cite{CMP06}

The performance of the virial series truncated after the fourth coefficient can be inspected by comparison with the exact equation of state.\cite{KP93,P20} The conventional truncated series from Eq.~\eqref{eq:virial_density} would be $Z\to Z_{\text{tr}}(\lambda)\equiv 1+B_2\lambda+B_3\lambda^2+B_4\lambda^3$. Alternatively, with the same amount of information, one can truncate the series at the level of Eq.~\eqref{eq:2.9b} to obtain $Z\to Z_{\text{tr}}'(p)\equiv 1+B_2 p+B_3'p^2+B_4' p^3$, where the density dependence of the compressibility factor is defined in parametric form ($p$ being the parameter) by the pair $Z=Z_{\text{tr}}'(p)$ and $\lambda=p/Z_{\text{tr}}'(p)$. As Fig.~\ref{fig:z_truncated} shows, the truncated series  $Z_{\text{tr}}(\lambda)$ is reliable only for $\lambda\lesssim 0.4$, whereas the truncated series $Z_{\text{tr}}'(p)$ is very accurate even at $\lambda\approx 1$, especially for small pore widths. This is not surprising given the fact that the exact equation of state for hard rods is $Z=1+B_2 p$ (with $B_2=1$).\cite{T36}
On the other hand, neither $Z_{\text{tr}}(\lambda)$ nor $Z_{\text{tr}}'(p)$ capture the divergence of pressure in the limit $\lambda\to\lambda_\cl$ discussed in Sec.~\ref{sec2D}.

Before turning to the high-pressure limit in Sec.~\ref{sec2D}, let us draw two relevant points from the analysis in this section. First, if for a given confined fluid with an unknown exact solution  one needs to resort to the virial coefficients (either analytically or numerically), then the standard irreducible diagrams do not provide the right answer. Instead, one would need to go back to the derivation steps\cite{S16} and include the reducible diagrams as well, which fail to cancel if the translational invariance is broken down. Second, if the first few virial coefficients are known and a truncated equation of state is employed as an approximation, the recommendation is to employ the pressure representation,\cite{MSRH11} Eq.~\eqref{eq:2.9b}, rather than the density representation, Eq.~\eqref{eq:virial_density}.

\subsection{\label{sec:system_highpressure} High-pressure behavior}
\label{sec2D}

Solving numerically the eigenvalue problem in Eq.~\eqref{eq:eigenvalue_equation} becomes increasingly more difficult as  pressure grows and the system approaches the close-packing limit. It is, therefore, of interest to study  analytically the limit $p\to\infty$ (or, equivalently, $\lambda\to\lambda_\cl$) in order to understand the full behavior of the system.

In this high-pressure limit, particles accumulate more and more near the walls, which means that $\phi(y)$ becomes non-zero only in two symmetric layers near $y=\pm \frac{\epsilon}{2}$. As a consequence,  the eigenfunction $\phi(y)$ and the eigenvalue $\ell$ for high values of $p$ adopt the forms (see Appendix \ref{app:C} for details)
\begin{subequations}
\label{eq:high-p}
\begin{equation}
\label{eq:phi_high}
    \phi(y) \to \frac{1}{\sqrt{\mathcal{N}}}\left[ \phi_+(y)+\phi_-(y)\right],
    \quad \phi_\pm(y)\equiv e^{-a(y\pm\frac{\epsilon}{2})p},
\end{equation}
\begin{equation}
\label{eq:l_high}
    \ell \to \frac{a(\epsilon)}{2\epsilon p}e^{-a(\epsilon)p}.
\end{equation}
\end{subequations}
In Eq.~\eqref{eq:phi_high}, the normalization constant is
\beq
\label{eq:2.22a}
\mathcal{N}\to   \frac{a(\epsilon)}{\epsilon p}e^{-2a(\epsilon)p}.
\eeq
It should be noted that, for high $p$,  $\phi_\pm(y)$ is practically nonzero only inside a region of width of the order of $a(\epsilon)/\epsilon p$, adjacent to the wall at $y=\pm \frac{\epsilon}{2}$.

\begin{table}
\caption{\label{table:2}Comparison between exact and MC values\cite{M20} of $Z$ and the high-pressure asymptotic form, Eq.~\eqref{eq:Z_high}.}
\begin{ruledtabular}
\begin{tabular}{ccccc}
$\epsilon$&$p$&$Z_{\text{exact}}$ &$Z_{\text{MC}}$&$2+a(\epsilon)p$\\
\hline
$0.4$&$12$&$12.774$&$12.774$&$12.998$\\
&$120$&$112.04$&$112.03$&$111.98$\\
$0.8$&$12$&$9.6547$&$9.6548$&$9.2000$\\
&$120$&$74.017$&$74.016$&$74.000$\\
\end{tabular}
\end{ruledtabular}
\end{table}

As proved in Appendix \ref{app:C}, the high-pressure compressibility factor becomes
\beq
\label{eq:Z_high}
Z\to 2+a(\epsilon)p.
\eeq
Table \ref{table:2} shows that  exact and MC simulation data\cite{M20} confirm the validity of Eq.~\eqref{eq:Z_high} as pressure increases.
Recalling that $\lambda_\cl=1/a(\epsilon)$, Eq.~\eqref{eq:Z_high} can be recast as
\begin{equation}\label{eq:z_residue}
Z\to\frac{2}{1-\lambda/\lambda_{\text{cp}}}.
\end{equation}

\begin{figure}
    \centering
    \includegraphics[width=0.8\columnwidth]{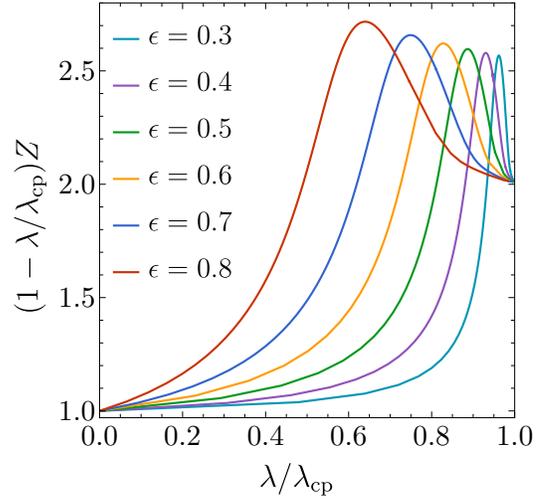}
    \caption{Normalized compressibility factor $(1-\lambda/\lambda_\cl)Z$ vs $\lambda/\lambda_{\text{cp}}$ for (from right to left) $\epsilon=0.3$, $0.4$, \ldots, $0.8$.}
    \label{fig:z_residue}
\end{figure}

Equation~\eqref{eq:z_residue} embodies two important features of the high-pressure asymptotic behavior of the compressibility factor. First, $Z$ presents a simple pole at $\lambda=\lambda_{\text{cp}}$, as expected. Second, the residue of the pole is not $1$ (as happens in the hard-rod Tonks gas,\cite{T36}) but $2$.
These two features are made quite apparent in  Fig.~\ref{fig:z_residue}, where the exact \emph{normalized} compressibility factor $(1-\lambda/\lambda_\cl)Z$ is plotted as a function of the scaled density  $\lambda/\lambda_{\text{cp}}$ for several values of $\epsilon$.
It can be observed that the normalized quantity  $(1-\lambda/\lambda_\cl)Z$ starts growing with density, then reaches a peak at a certain value $\lambda_{\text{peak}}$, and subsequently decays toward its asymptotic value $2$. We have checked that $\lambda_{\text{peak}}$ is slightly higher than $1$ for any $\epsilon$, namely, $\lambda_{\text{peak}}\simeq 1+\frac{1}{10}\epsilon^2$. Thus, in the region of small pore width, one has $1-\lambda/\lambda_\cl\approx\frac{2}{5}\epsilon^2$.
It is then obvious that the limiting value $(1-\lambda/\lambda_\cl)Z\to 2$ requires linear densities closer and closer to $\lambda_\cl$ as $\epsilon$ decreases. In fact, in the Tonks gas, $\lambda_\cl=1$ and $Z=1/(1-\lambda)$ for any density. This shows that the limits $p\to\infty$ and $\epsilon\to 0$ do not commute and that a significant difference  between 1D and Q1D systems exists, one of the additional key differences being the existence of a transverse pressure in the latter systems.\cite{P20}

\section{Approximate Equations of State}\label{sec:approximations}

In order to obtain the exact equilibrium properties of the confined hard-disk system, one  needs to  solve Eq.~\eqref{eq:eigenvalue_equation}, which, however, does not seem to have any known analytical solution, so that one must resort to numerical methods.\cite{KP93} Some authors have proposed to  simplify the model by replacing $a(s)$ by its linear approximation, Eq.~\eqref{eq:as_series},\cite{VBG11}  or by means of  fitting parameters.\cite{KMP04}

We propose here an alternative approach  that does not rely on solving Eq.~\eqref{eq:eigenvalue_equation} or using any fitting parameters, but instead benefits from the study of the physical properties in the low- and high-pressure limits.
For this purpose, it is convenient to consider the equation of state as written in Eq.~\eqref{eq:z_exact_02}, where the eigenvalue $\ell$ does not appear explicitly  and, therefore, $\phi(y)$ does not need to be normalized.

In the following discussion, two different analytic approximations for $\phi(y)$ will be proposed and discussed, which will be referred to as the \emph{uniform-profile} approximation (UPA) and the \emph{exponential-profile} approximation (EPA).

\subsection{Uniform-profile approximation}

Under low-pressure (and, therefore, low-density) conditions, particles barely interact with one another and are then allowed to move almost freely around the available space. This setup yields a nearly uniform density profile along the transverse direction. In the limit  $p \rightarrow 0$, this density profile is exactly constant, as shown in Appendix~\ref{app:virial_series_math}.

Based on this behavior, we construct here the UPA by taking $\phi(y)=\text{const}$ not only for $p\to 0$ but for any value of $p$. As we will see, despite its crudeness, the UPA can provide reasonable results, except for very high pressures and/or wide pores. Under this approximation, Eq.~\eqref{eq:z_exact_02}  yields
\begin{equation}\label{eq:z_basic_00}
      Z_{\mathrm{UPA}} = 1 + p \,\frac{\int\mathrm{d}y_1\int\mathrm{d}y_2 e^{-a(y_1-y_2)p}a(y_1-y_2)}{\int\mathrm{d}y_1\int\mathrm{d}y_2 e^{-a(y_1-y_2)p}}.
\end{equation}
Then, by setting $s=y_1 - y_2$ and using the mathematical identity
\begin{equation}\label{eq:basic_integration_change}
    \int\mathrm{d}y_1\int\mathrm{d}y_2\,F(y_2-y_1) = \int_{0}^\epsilon \mathrm{d}s \left[ F(s) + F(-s)\right](\epsilon-s),
\end{equation}
Eq.~\eqref{eq:z_basic_00} can be simplified as
\begin{equation}
\label{Z_UPA}
    Z_{\mathrm{UPA}} = 1+p\frac{\int_0^{\epsilon} \mathrm{d}s \, a(s)(\epsilon-s)e^{-a(s)p}}{\int_0^{\epsilon} \mathrm{d}s \, (\epsilon-s)e^{-a(s)p}}.
\end{equation}

Expanding in powers of $p$ in both the numerator and the denominator of Eq.~\eqref{Z_UPA}, it is not difficult to obtain the virial coefficients in this UPA. As expected, the second virial coefficient $B_2$ is recovered, while the higher-order virial coefficients are approximate. In particular,
\begin{subequations}
\bal
B_{3,\text{UPA}}'=&-\left(1-B_2^2-\frac{\epsilon^2}{6}\right)\nn
=&-\frac{7 \epsilon^4}{720}\left(1 + \frac{31 \epsilon^2}{98} + \frac{261 \epsilon^4}{1960}+\cdots\right),
\eal
\bal
B_{4,\text{UPA}}'=&B_2^3-B_2\left(\frac{9}{8}-\frac{\epsilon^2}{4}\right)+\frac{1-(1-\epsilon^2)^{5/2}}{20\epsilon^2}\nn
=&-\frac{11 \epsilon^6}{15120}\left(1 + \frac{543 \epsilon^2}{880} + \frac{14259 \epsilon^4}{38720}+\cdots\right).
\eal
\end{subequations}

In the opposite high-pressure limit, an analysis similar to that described in Appendix \ref{app:C} yields $Z_{\text{UPA}}\to 3+a(\epsilon)p$, which implies
\begin{equation}
Z_{\text{UPA}}\to\frac{3}{1-\lambda/\lambda_{\text{cp}}}.
\end{equation}
Thus, the UPA predicts the right pole at $\lambda=\lambda_{\text{cp}}$ but overestimates the residue by $50\%$.

\subsection{Exponential-profile approximation}

On a different vein, the EPA is constructed by taking  $\phi(y)$ in the same functional form as in the limit $p\to\infty$, Eq.~\eqref{eq:phi_high}, except that now $p$ is assumed to be arbitrary. It should be noted that in the EPA, the transverse density decays exponentially near the walls at $y=\pm \frac{\epsilon}{2}$, hence the name of the approximation. Within this approximation, the compressibility factor becomes
\begin{widetext}
\begin{equation}\label{eq:z_EPA}
      Z_{\text{EPA}} = 1 + p \frac{\int\mathrm{d}y_1\int\mathrm{d}y_2 \, e^{-a(y_1-y_2)p}a(y_1-y_2)\phi_+(y_1)\left[\phi_+(y_2)+\phi_-(y_2)\right]}{\int\mathrm{d}y_1\int\mathrm{d}y_2 \, e^{-a(y_1-y_2)p}\phi_+(y_1)\left[\phi_+(y_2)+\phi_-(y_2)\right]},
\end{equation}
\end{widetext}
where we have used the symmetry property $\phi_-(y)=\phi_+(-y)$.

Even though the EPA is inspired by the exact high-pressure behavior, Eq.~\eqref{eq:z_EPA} makes sense even for low $p$. In fact, since $\lim_{p\to 0}\phi_\pm(y)=1$, both the EPA and the UPA yield the exact second virial coefficient.
Expanding the numerator and the denominator of Eq.~\eqref{eq:z_EPA} in powers of $p$, and after some algebra, one finds
\begin{subequations}
\bal
B_{3,\text{EPA}}'=&-\left[1-\frac{\epsilon^2}{6}-2B_2^2-2B_2\frac{1-(1-\epsilon^2)^{3/2}}{3\epsilon^2}+2U_2\right]\nn
=&-\frac{\epsilon^4}{80} \left(1 + \frac{8 \epsilon^2}{21} + \frac{58 \epsilon^4}{315}+\cdots\right),
\eal
\bal
B_{4,\text{EPA}}'=&\frac{15}{4}B_2^3
-B_2\left(\frac{4+2\epsilon^2+\epsilon^4}{4\epsilon^2}+6U_2\right)+ \frac{2}{3}+U_3\nn
&+\left(7B_2^2-\frac{1}{3}-4U_2+\frac{2}{\epsilon^2}B_2\right)\frac{1-(1-\epsilon^2)^{3/2}}{3\epsilon^2}\nn
=&-\frac{\epsilon^6}{504} \left(1 + \frac{279 \epsilon^2}{400} + \frac{2041 \epsilon^4}{4400}+\cdots\right),
\eal
\end{subequations}
where
\begin{subequations}
\bal
\label{eq:U2}
U_2\equiv&\frac{1}{\epsilon}\int\mathrm{d}y\,\psi_1(y)a\left(y+\frac{\epsilon}{2}\right)\nn
=&1 - \frac{\epsilon^2}{4} - \frac{13 \epsilon^4}{720} - \frac{23 \epsilon^6}{3360}+\cdots,
\eal
\bal
U_3\equiv&\frac{1}{2\epsilon^2}\int\mathrm{d}y_1\int\mathrm{d}y_2\,a(y_1-y_2)a\left(y_1+\frac{\epsilon}{2}\right)\nn
&\times\left[a\left(y_2+\frac{\epsilon}{2}\right)+a\left(y_2-\frac{\epsilon}{2}\right)\right]\nn
=&1 - \frac{5 \epsilon^2}{12} - \frac{17 \epsilon^6}{2880}+\cdots.
\eal
\end{subequations}
In Eq.~\eqref{eq:U2}, the function $\psi_1(y)$ is defined by Eq.~\eqref{eq:virial_int_00}

\section{Assessment of the Uniform-Profile and Exponential-Profile Approximations}
\label{sec:4}

The main idea behind both the UPA and EPA consists in  replacing the actual eigenfunction $\phi(y)$ in the numerator and denominator integrals of Eq.~\eqref{eq:z_exact_02} by simple approximate functions. It is now convenient to study how well the system is described by these two approximations, as well as their range of validity.
For that purpose, we analyze, in this section, several properties of the system, comparing the proposed approximations with the numerical solution corresponding to the exact description presented in Sec.~\ref{sec:solution}.
Some technical details about our numerical solution of the eigenvalue problem, Eq.~\eqref{eq:eigenvalue_equation}, and the numerical evaluation of the compressibility factor from Eqs.~\eqref{eq:z_exact_01}, \eqref{Z_UPA}, and \eqref{eq:z_EPA} are given in Appendix \ref{app:numerical_methods}.

\subsection{Transverse density  profiles}
\begin{figure}
\includegraphics[height=0.6\columnwidth]{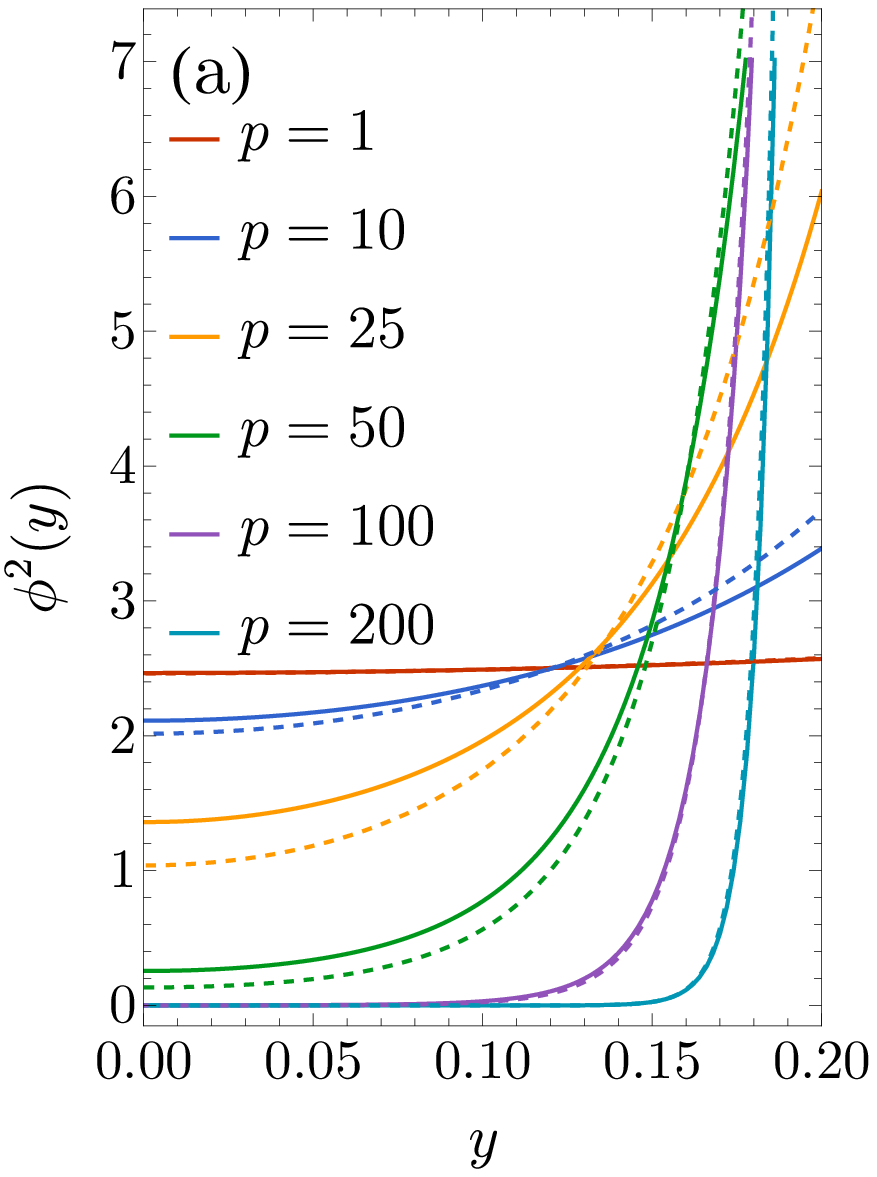}
\includegraphics[height=0.6\columnwidth]{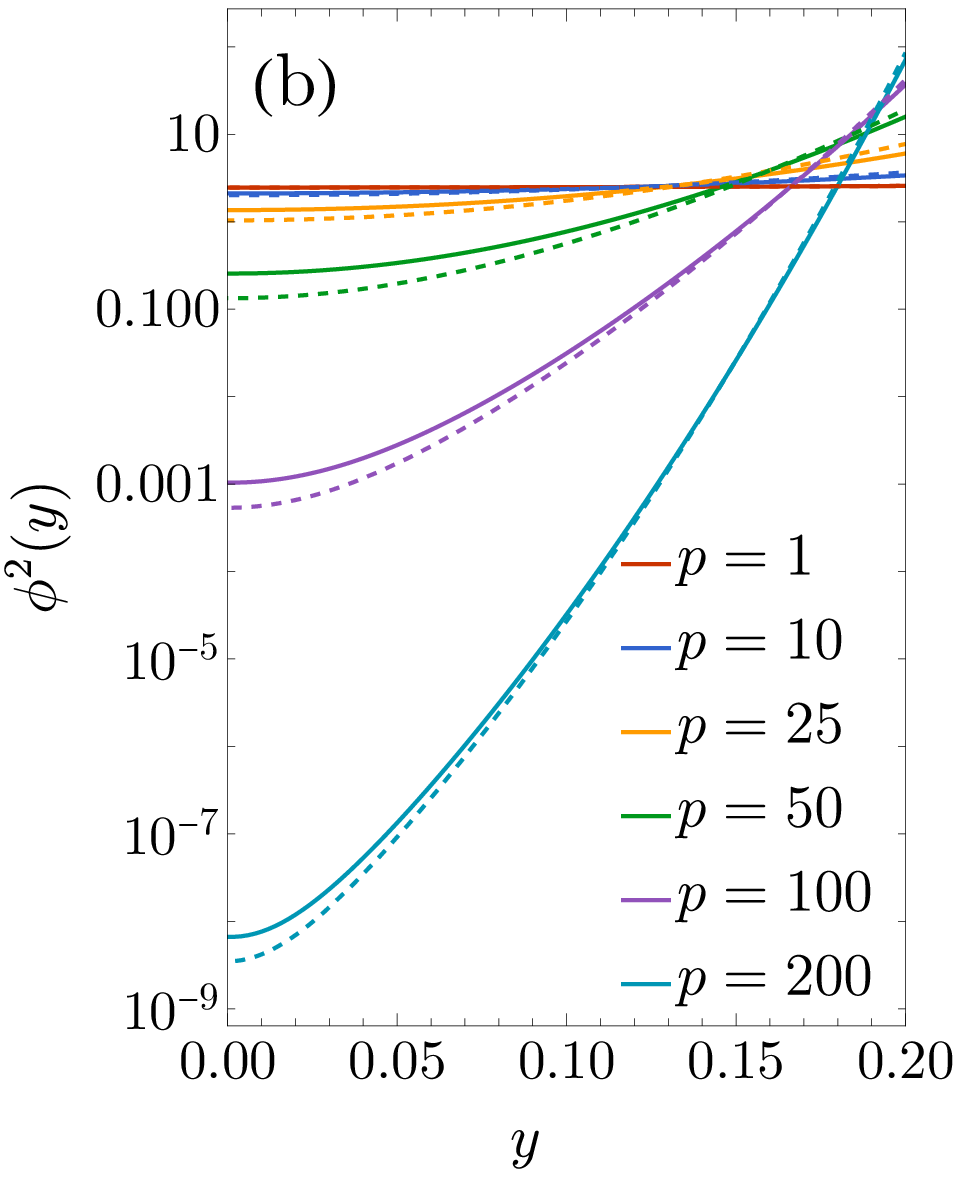}
\caption{Plot of the transverse density profile $\phi^2(y)$ as obtained from the numerical solution of Eq.~\eqref{eq:eigenvalue_equation} (solid lines) and as given by the EPA, Eq.~\eqref{eq:phi_high} (dashed lines), for $\epsilon=0.4$ and several values of $p$. In panels (a) and (b), the vertical axis is in normal and logarithmic scale, respectively. It should be noted that, due to symmetry, only the region $0\leq y\leq\frac{\epsilon}{2}$ is considered.}
     \label{fig:profile_advanced}
\end{figure}

Figure~\ref{fig:profile_advanced} shows a comparison between the exact (numerical) transverse  density profile coming from Eq.~\eqref{eq:eigenvalue_equation} and the  EPA analytical profile, Eq.~\eqref{eq:phi_high}, for  $\epsilon=0.4$ and some representative values of $p$.
It should be noted that here the normalization constant $\mathcal{N}$ is not given by Eq.~\eqref{eq:2.22a} but is instead obtained by requiring fulfillment of Eq.~\eqref{eq:eigenvalue_normalization}. Although this normalization constant is  not needed in Eq.~\eqref{eq:z_EPA}, it is used  in Fig.~\ref{fig:profile_advanced}.

We observe that, even though the EPA was based  on the exact high-pressure limit behavior,  a  good agreement with the numerical solution is reached for all pressure ranges, including the low-pressure regime, where  the solution $\phi \approx \text{const}$ is recovered. In fact, we find that the worst agreement is centered around the medium pressure regime. Similar results can also be found for other values of the width parameter $\epsilon$.

\subsection{Virial coefficients}

\begin{figure}
\includegraphics[height=0.57\columnwidth]{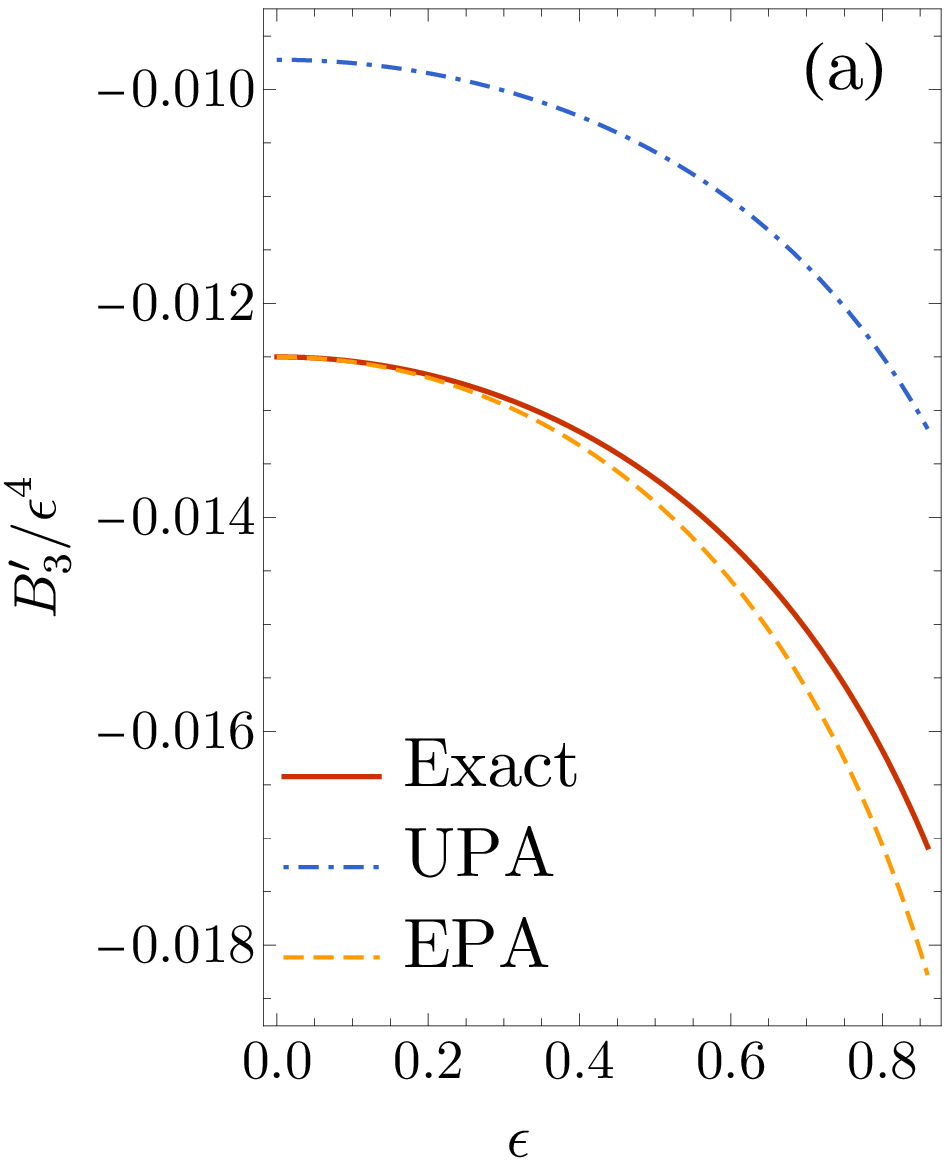}
\includegraphics[height=0.57\columnwidth]{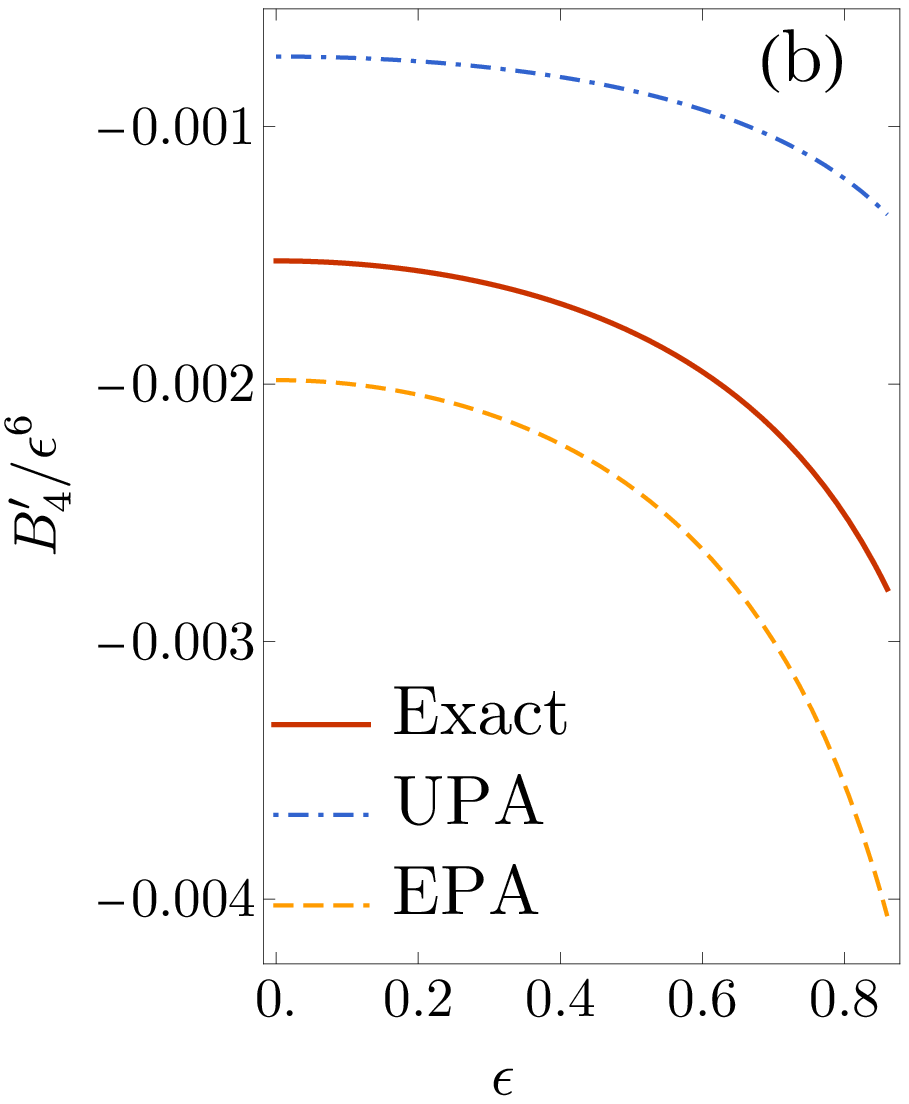}
     \caption{Plot of (a) $B_3'/\epsilon^4$ and (b) $B_4'/\epsilon^6$ as functions of the excess pore width $\epsilon$. The solid, dashed, and dashed-dotted lines correspond to the exact, EPA, and UPA results, respectively.}
     \label{fig:B3p&B4p}
\end{figure}

Figure~\ref{fig:B3p&B4p} compares the exact and approximate values of $B_3'/\epsilon^4$ and $B_4'/\epsilon^6$. As can be observed, the EPA predictions are more accurate than the UPA ones. On the other hand, since $B_3'$ and $B_4'$ are rather small,  the conventional virial coefficients $B_3$ and $B_4$  are dominated by $B_2^2$ and $B_2^3$, respectively [see Eq.~\eqref{eq:2.10}]. Thus,  the impact on $B_3$ and $B_4$ of the deviations observed in Fig.~\ref{fig:B3p&B4p} is very small. At the maximum excess width, $\epsilon_{\max}=\sqrt{3}/2\simeq 0.866$, we have observed that the relative deviations in $B_3$ are approximately $0.3\%$ (UPA) and $-0.03\%$ (EPA), while, in the case of $B_4$, they are approximately $-0.5\%$ (UPA) and $0.04\%$ (EPA).

\subsection{Equation of state}

The equation of state involves performing the integrals in Eq.~\eqref{eq:z_exact_02} once the density profiles (either exact or approximate) are known.

\begin{figure}
    \centering
    \includegraphics[width=0.8\columnwidth]{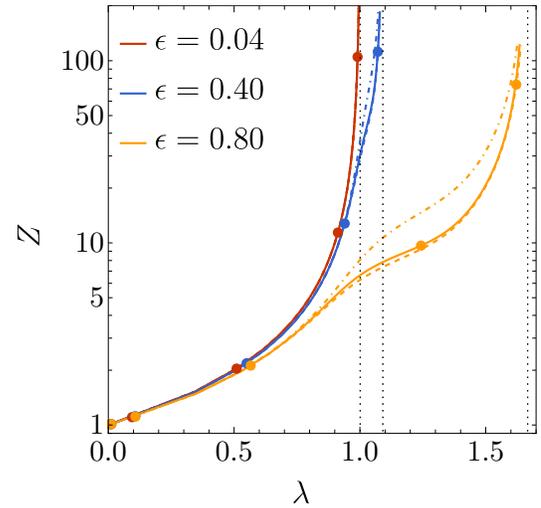}
    \caption{Compressibility factor as a function of the longitudinal density $\lambda$ for different values of the excess pore width $\epsilon$. The circles represent MC data,\cite{M20} while the solid, dashed, and dashed-dotted lines correspond to exact, EPA, and UPA results, respectively. The vertical lines denote the locations of $\lambda_\cl$.}
    \label{fig:comparison_z}
\end{figure}

Figure~\ref{fig:comparison_z} depicts the comparison between the two proposed approximations and the results coming from both the numerical evaluation of the exact solution for the Q1D fluid and independently calculated MC simulations for the original confined 2D system.\cite{M20} It shows a good agreement with the UPA under low-pressure and/or narrow-pore conditions, and a very good agreement with the EPA for practically all ranges of pressure and pore sizes. In the case of the EPA, the results disagree visibly from the exact solution only within a small region of medium pressures for  large values of the pore size.
It is interesting to note that the compressibility factor, especially  with an excess pore width $\epsilon=0.80$, presents two inflection points, a feature captured even by the UPA. Although the system lacks a true phase transition, those  two inflection points can be seen as  precursors of the phase transition in genuine 2D systems.\cite{BK11,GM14}

\begin{figure}
     \centering
         \includegraphics[width=0.8\columnwidth]{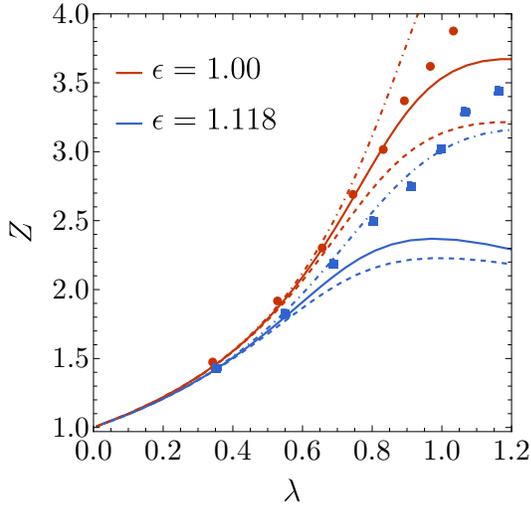}
         \caption{Compressibility factor as a function of the longitudinal density $\lambda$ for two values of $\epsilon$ beyond the nearest-neighbor condition: $\epsilon=1$ and $1.118$. The symbols represent MC data,\cite{KP93} while the solid, dashed, and dashed-dotted lines correspond to results from the solution of the eigenvalue problem, Eq.~\eqref{eq:eigenvalue_equation}, the EPA, and the UPA, respectively.}
         \label{fig:Z_morethanemax}
\end{figure}

Even though the transfer-matrix solution and our approximations were developed only for nearest-neighbor interactions (single-file condition), which precludes an excess width of the channel larger than $\epsilon_{\mathrm{max}}=\sqrt{3}/2$, it is also of interest to study how well the theoretical treatments behave when this limit is exceeded.\cite{KP93} In that case, the function $a(s)$ defined by Eq.~\eqref{eq:a(s)} must be supplemented as $a(s)=0$ if $s>1$.\cite{KP93}
A comparison with MC simulation data\cite{KP93} for $\epsilon=1$ and $1.118$ is shown in Fig.~\ref{fig:Z_morethanemax}. We observe that, as density or pressure increases, none of the three methods is accurate. Paradoxically, however, the UPA performs a reasonable job and is perhaps the most reliable approximation in the case $\epsilon=1.118$.

\subsection{Execution times}

\begin{figure}
     \centering
         \includegraphics[width=0.8\columnwidth]{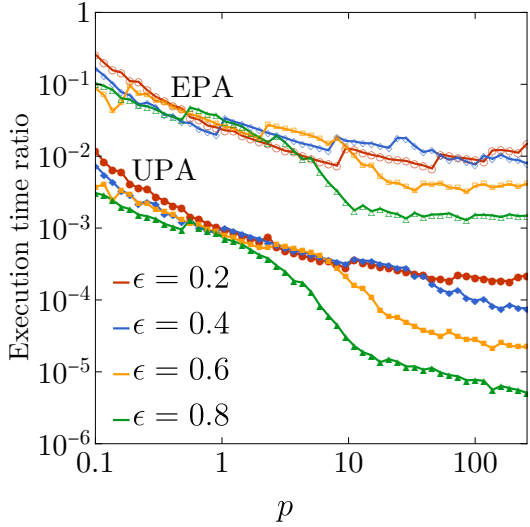}
         \caption{Wall time ratios between both approximations and the exact solution vs $p$ for some representative values of $\epsilon$. Closed and open symbols  represent the UPA and EPA values, respectively.
         Lines are guides to the eye.}
         \label{fig:cpu_times}
\end{figure}

In the transfer-matrix formalism,  as well as in our approximations,  the final computation of $Z$  must be performed numerically (see Appendix \ref{app:numerical_methods}). It is then  worth studying the different execution times (the so-called wall times\cite{WallTime}) in order to assess the cost of  using the exact solution against any of the two approximations proposed in this paper.

Figure~\ref{fig:cpu_times} shows the UPA-to-exact and EPA-to-exact wall time ratios. We clearly see that both approximations are much faster than the exact evaluation for all ranges of pressure and pore sizes, and that this wall time advantage increases with the increasing pressure and pore width. For the EPA, this is especially relevant in the case of large pore sizes and high pressures, where the performance of the EPA is excellent (see Fig.~\ref{fig:comparison_z}). In the case of the UPA, the gain in wall time is still very remarkable even for  small pore sizes and small or moderate pressures, where both the exact solution and the UPA practically yield the same results (see again Fig.~\ref{fig:comparison_z}).

\section{Concluding Remarks}
\label{sec: concl}
In this work, we have started from the exact equation of state of the single-file hard-disk confined fluid, as derived from the  transfer-matrix method.\cite{KP93} We showed that exactly the same result is also obtained by mapping the original system onto a 1D polydisperse mixture of \emph{non-additive} hard rods with a common chemical potential, in contrast to previous approximate mappings to hard-rod additive mixtures.\cite{PK92}

From the exact solution, we then explored the low-pressure regime by using  a perturbation scheme to obtain the exact third and fourth virial coefficients, which, to the best of our knowledge, were still unknown. The results differ from a recent alternative derivation\cite{M20} based on the standard irreducible diagrams, thus showing  that the conventional cancellation of the reducible diagrams does not hold for confined fluids, a fact usually overlooked in the literature.\cite{M14b,M15,M20}

The high-pressure regime, near the close-packing region, was also studied in order to get the asymptotic behavior of the equation of state, which is seen to present a simple pole at the close-packing linear density with a residue equal to $2$, in contrast to the residue equal to $1$ in the 1D Tonks gas.\cite{T36}

The study of the exact physical properties of the system allowed us to propose  two different approximations for the equation of state, namely, the UPA and the EPA. The first one has a much simpler form than the second one but its range of validity is restricted to narrow pores and/or low pressures, whereas the EPA is valid throughout the entire range of pore sizes and pressures, yielding results which are virtually indistinguishable from the exact solution, except in a small region of high pore sizes and intermediate pressures.

The usefulness and reliability of the approximations were tested for different quantities,  such as the transverse density profile, the virial coefficients, and the equation of state. In the case of the latter quantity, we also considered situations beyond the nearest-neighbor constraint $\epsilon\leq \epsilon_{\mathrm{max}}$ and even beyond the single-file condition $\epsilon\leq 1$.
Tests regarding execution times of the exact solution, on the one hand, and the two approximations, on the other hand, were performed in order to assess the practical convenience of using the approximate methods instead of the exact solution. Execution times for the approximate compressibility factors were found to be $10$--$10^3$ times and  $10^2$--$10^{5}$ times faster in the cases of the EPA and UPA, respectively.

We plan to exploit the 1D mapping to obtain the structural correlation functions of the confined hard-disk fluid. In addition, the extensions of the UPA and EPA for the hard-sphere fluid confined in a narrow cylindrical pore will be undertaken in the near future.

\acknowledgments
The authors acknowledge financial support from Grant No.~PID2020-112936GB-I00 funded by MCIN/AEI/10.13039/501100011033, and from Grant No.~IB20079 funded by Junta de Extremadura (Spain) and by ERDF
``A way of  making Europe.''
A.M.M. is grateful to the Spanish Ministerio de Ciencia e Innovaci\'on for a predoctoral fellowship PRE2021-097702.

\section*{AUTHOR DECLARATIONS}
\subsection*{Conflict of Interest}
The authors have no conflicts to disclose.
\subsection*{Author Contributions}
\textbf{Ana M. Montero}: Formal analysis (equal); Investigation (equal);
Methodology (equal); Software (lead); Writing -- original draft
(lead). \textbf{Andr\'es Santos}: Conceptualization (lead); Formal analysis
(equal); Funding acquisition (lead); Investigation (equal); Methodology
(equal); Supervision (lead); Writing -- original draft (supporting);
Writing -- review \& editing (lead).

\section*{Data availability}
The data that support the findings of this study are available from the corresponding author upon reasonable request.

\appendix
\section{Mapping onto a one-dimensional polydisperse mixture of non-additive hard rods}\label{app:mapping_onedimension}

When one focuses on the longitudinal properties, the original system under study can be mapped onto a 1D polydisperse hard-rod non-additive mixture, where the transverse coordinate $-\epsilon/2 \leq y \leq \epsilon/2$ of each disk plays the role of the dispersity parameter. Under this framework, two hard rods of different \emph{species} $y$ and $y'$ interact with an effective hard-core distance $a(y-y') = \sqrt{1-(y-y')^2}$. The equation of state of such a system can, in principle, be obtained exactly.

Let us consider first a discrete $M$-component mixture, where each 1D component $i$ represents  disks with a transverse coordinate
\begin{equation}
\label{eq:delta_y}
y_i=-\frac{\epsilon}{2}+(i-1)\delta y,\quad i=1,2,\ldots,M,\quad \delta y\equiv \frac{\epsilon}{M-1}.
\end{equation}
In that case, the hard-core distance between two rods of species $i$ and $j$ is
\begin{equation}
a_{ij}\equiv a(y_i-y_j)=\sqrt{1-\left[(i-j)\delta y\right]^2}.
\end{equation}
It is worth noting that $a_{ii}=1$ but $a_{ij}<1$ if $i\neq j$ so that the hard-rod mixture is negatively non-additive.

From the classical theory of liquids,\cite{S16} one can derive the equation of state as given by
\begin{equation}
\label{eq:A3}
    -\frac{1}{\lambda}=\sum_{i,j}\sqrt{x_i x_j} A_iA_j \Omega'_{ij}(p), \quad x_i=\frac{N_i}{N},
\end{equation}
where  $N_i$ is the number of particles of species $i$, $\Omega_{ij}'(p) =- \Omega_{ij}(p)(a_{ij}+1/p)$ is the derivative of $\Omega_{ij}(p)=e^{-a_{ij} p}/p$, and the coefficients $A_{i}$ are related to the mole fractions by
    \beq
    \label{eq:map_condition02}
    \sum_{j}\sqrt{x_j} A_{i}A_j \Omega_{ij}(p)=\sqrt{x_i}.
\eeq
From Eq.~\eqref{eq:map_condition02}, one has
\beq
\label{eq:A5}
\sum_{i,j}\sqrt{x_ix_j} A_{i}A_j \Omega_{ij}(p)=1.
\eeq
As a consequence, Eq.~\eqref{eq:A3} can be rewritten as
\beq
\label{eq:A6}
Z=1+\sum_{i,j}\sqrt{x_i x_j} A_{i}A_j a_{ij}e^{-a_{ij}p}.
\eeq

In an ordinary 1D mixture, the mole fractions $\{x_i\}$ are independent variables and the coefficients $A_{i}$ must be found from Eq.~\eqref{eq:map_condition02} as functions of the mole fractions and the pressure. In our case, however, since the original Q1D system is made of identical disks, the mole fractions of the mapped 1D fluid are constrained by the condition that the chemical potential of all the components  must be the same. It can be checked that this condition implies that all $A_{ij}=A$ are equal. In that case, Eqs.~\eqref{eq:map_condition02} and \eqref{eq:A6} become
\begin{subequations}
\label{eq:A7}
\beq
\label{eq:A7a}
\sum_{j}\sqrt{x_j}  e^{-a_{ij}p}=\frac{p}{A^2}\sqrt{x_i},
\eeq
\beq
\label{eq:A7b}
Z=1+A^2\sum_{i,j}\sqrt{x_i x_j} a_{ij}e^{-a_{ij}p}.
\eeq
\end{subequations}
Finally, identifying ${x_i}\to\phi^2(y_i)\delta y$ and $A^2\to (p/\ell)\delta y$, and then taking the continuum limit ($M\to\infty$), where $\delta y\sum_i\to \int\mathrm{d}y$, one obtains Eqs.~\eqref{eq:eigenvalue_equation} and \eqref{eq:z_exact_01} from Eqs.\ \eqref{eq:A7a} and \eqref{eq:A7b}, respectively.

The exact mapping described here differs from the approximate one in Ref.~\onlinecite{PK92}, since in the latter reference, each rod has a different size and the mixture is assumed to be additive.

\section{Virial series expansion}\label{app:virial_series_math}
Let us start by listing here some integrals involving the function $a(s)$ that will be useful later on,
\begin{subequations}
\beq
\label{eq:virial_int_00}
\psi_1(y_1)\equiv \frac{1}{\epsilon}
\int\mathrm{d}y_2\, a(y_1-y_2)=\frac{1}{2\epsilon}\left[\bar{\psi}(y_1)+\bar{\psi}(-y_1)\right],
\eeq
\beq
\bar{\psi}(y)\equiv\left(\frac{\epsilon}{2}+ y\right)\sqrt{1-\left(\frac{\epsilon}{2}+ y\right)^2}+\sin^{-1}\left(\frac{\epsilon}{2}+ y\right),
\eeq
\begin{equation}
\label{eq:virial_int_01}
\frac{1}{\epsilon}\int\mathrm{d}y_2\, a^2(y_1-y_2)=1-\frac{\epsilon^2}{12}-y_1^2,
\end{equation}
\beq
\label{eq:virial_int_02}
\frac{1}{\epsilon^2}\int\mathrm{d}y_1\int\mathrm{d}y_2\,a(y_1-y_2)=\frac{1}{\epsilon}\int\mathrm{d}y\,\psi_1(y)
= B_2,
\eeq
\begin{equation}
\label{eq:virial_int_03}
\frac{1}{\epsilon^2}\int\mathrm{d}y_1\int\mathrm{d}y_2\,a^2(y_1-y_2)=1-\frac{\epsilon^2}{6},
\end{equation}
\bal
\label{eq:Q}
Q\equiv &\frac{1}{\epsilon^2}\int\mathrm{d}y_1\int\mathrm{d}y_2\,a^3(y_1-y_2)\nn
=&\frac{3}{4}B_2-\frac{(1-\epsilon^2)^{5/2}-1}{10\epsilon^2},
\eal
\bal
\label{eq:S}
S\equiv &\frac{1}{\epsilon}\int\mathrm{d}y\,\psi_1(y)y^2\nn
=&\left(\frac{1}{8}+\frac{\epsilon^2}{12}\right)B_2+\frac{(1-\epsilon^2)^{5/2}-1-20\epsilon^2}{180\epsilon^2}.
\eal
\end{subequations}
In Eqs.~\eqref{eq:virial_int_02}, \eqref{eq:Q}, and  \eqref{eq:S}, $B_2$ is given by Eq.~\eqref{eq:B2}.

Now we  proceed to the derivation of $\phi_0(y)$, $\phi_1(y)$, $\phi_2(y)$, $\ell_0$, $\ell_1$, and $\ell_2$. Insertion of Eq.~\eqref{eq:virial_series} into Eqs.~\eqref{eq:eigenvalue_equation} and \eqref{eq:eigenvalue_normalization} yields
\begin{subequations}
\begin{equation}
\label{eq:virial_sol_00}
\int\mathrm{d}y_2\,\phi_0(y_2)=\ell_0\phi_0(y_1),
\end{equation}
\begin{equation}
\label{eq:virial_sol_01}
\int\mathrm{d}y_2\,\left[\phi_1(y_2)-a(y_1-y_2)\phi_0(y_2)\right]=\ell_0\phi_1(y_1)+\ell_1\phi_0(y_1),
\end{equation}
\begin{eqnarray}
\label{eq:virial_sol_02}
\int\mathrm{d}y_2\,&&\left[\phi_2(y_2)-a(y_1-y_2)\phi_1(y_2)+\frac{1}{2}a^2(y_1-y_2)\phi_0(y_2)\right]\nonumber\\
&&=\ell_0\phi_2(y_1)+\ell_1\phi_1(y_1)+\ell_2\phi_0(y_1),
\end{eqnarray}
\end{subequations}
\begin{subequations}
\begin{equation}
\label{eq:virial_norm_00}
\int\mathrm{d}y\,\phi_0^2(y)=1,
\end{equation}
\begin{equation}
\label{eq:virial_norm_01}
\int\mathrm{d}y\,\phi_0(y)\phi_1(y)=0,
\end{equation}
\begin{equation}
\label{eq:virial_norm_02}
\int\mathrm{d}y\,\left[\phi_1^2(y)+2\phi_0(y)\phi_2(y)\right]=0.
\end{equation}
\end{subequations}
Equation~\eqref{eq:virial_sol_00} implies that $\phi_0(y)$ is a constant, and using the normalization condition, Eq.~\eqref{eq:virial_norm_00}, we obtain
\begin{equation}
\label{eq:B4}
    \phi_0(y) = \frac{1}{\sqrt{\epsilon}},\quad     \ell_0 = \epsilon.
\end{equation}
Next, we note from  Eq.~\eqref{eq:virial_sol_01} that
\begin{equation}
    \phi_1(y) = -\frac{1}{\sqrt{\epsilon}}\left[ \psi_1(y)- \alpha_1\right],
    \quad  \alpha_1\equiv \frac{1}{\sqrt{\epsilon}}\int\mathrm{d}y\,\phi_1(y)-\frac{\ell_1}{\epsilon}.
\end{equation}
From the definition of $\alpha_1$ we obtain $\ell_1=-\epsilon B_2$, while use of Eq.~\eqref{eq:virial_norm_01} implies that $\alpha_1=B_2$. Therefore,
\begin{equation}
\label{eq:B6}
    \phi_1(y)=-\frac{1}{\sqrt{\epsilon}}\left[\psi_1(y)- B_2\right],\quad
    \ell_1=-\epsilon B_2.
\end{equation}

Finally, we evaluate $\phi_2(y)$ and $\ell_2$. Equation~\eqref{eq:virial_sol_02} gives
\begin{equation}
\label{eq:B7}
    \phi_2(y) = \frac{1}{\sqrt{\epsilon}}\left[\psi_2(y)-2B_2\psi_1(y) - \frac{1}{2}y^2 + \alpha_2 \right],
\end{equation}
where
\begin{subequations}
\begin{equation}
\label{eq:psi2}
    \psi_2(y_1) \equiv\frac{1}{\epsilon}\int\mathrm{d}y_2 \, a(y_1-y_2)\psi_1(y_2),
\end{equation}
\begin{equation}
   \alpha_2\equiv \frac{1}{\sqrt{\epsilon}}\int\mathrm{d}y\,\phi_2(y)+\frac{1}{2}\left(1-\frac{\epsilon^2}{12}\right)+ B_2^2-\frac{\ell_2}{\epsilon}.
\end{equation}
\end{subequations}
The definition of $\alpha_2$ yields
\begin{equation}
\label{eq:B9}
    \ell_2= \epsilon\left(\frac{1}{2}+W_2-B_2^2-\frac{\epsilon^2}{12} \right),
\end{equation}
where
\bal
\label{eq:W_2}
    W_2 \equiv &\frac{1}{\epsilon}\int\mathrm{d}y \, \psi_2(y)=\frac{1}{\epsilon}\int\mathrm{d}y \, \psi_1^2(y)\nn
    =&1 - \frac{\epsilon^2}{6} - \frac{\epsilon^4}{120} - \frac{13 \epsilon^6}{5040}+\cdots .
\eal
Using now the normalization condition in Eq.~\eqref{eq:virial_norm_02}, one also obtains
\begin{equation}
\label{eq:B11}
    \alpha_2=\frac{5}{2}B_2^2-\frac{3}{2}W_2+\frac{\epsilon^2}{24}.
\end{equation}
It should be noted that the function $\psi_2(y)$ and the constant $W_2$ defined by Eqs.~\eqref{eq:psi2} and \eqref{eq:W_2}, respectively, must be obtained numerically.
It can easily be checked that Eqs.~\eqref{eq:B4}, \eqref{eq:B6}, \eqref{eq:B7}, \eqref{eq:B9}, and \eqref{eq:B11} are consistent with Eq.~\eqref{eq:2.5}.

Once we have determined $\{\phi_n\}$ and $\{\ell_n\}$ for $n=0,1,2$, we can expand the integral $I$, as defined by Eq.~\eqref{eq:I}, resulting in $I_0=-\ell_1$ and $I_1=-2\ell_2$, in agreement with Eq.~\eqref{eq:In}.
Furthermore, the determination of $I_2$ allows one to obtain $\ell_3=-I_2/3$ as
\beq
\ell_3=-\epsilon\left[W_3+B_2\left(2B_2^2-3W_2+\frac{\epsilon^2}{12}\right)+\frac{Q}{6}-S\right],
\eeq
where
\bal
\label{eq:W_3}
    W_3 \equiv &\frac{1}{\epsilon}\int\mathrm{d}y \, \psi_1(y)\psi_2(y)\nn
    =&\frac{1}{\epsilon^2}\int\mathrm{d}y_1\int\mathrm{d}y_2 \, a(y_1-y_2)\psi_1(y_1)\psi_1(y_2)\nn
    =&1 - \frac{\epsilon^2}{4} - \frac{\epsilon^4}{720} - \frac{71 \epsilon^6}{30240}+\cdots.
\eal

\section{Limit $p\to\infty$}
\label{app:C}

Here, we prove Eqs.~\eqref{eq:high-p} and \eqref{eq:Z_high} in the limit $p\to\infty$.
Let us first obtain the normalization constant $\mathcal{N}$ from Eq.~\eqref{eq:phi_high}:
\bal
\label{eq:2.20}
\mathcal{N}=&\int \mathrm{d}y\,\left[e^{-2a(y+\frac{\epsilon}{2})p}+e^{-2a(y-\frac{\epsilon}{2})p}\right]\nn
=&2\int_0^\epsilon \mathrm{d}s\, e^{-2a(s)p},
\eal
where we have taken into account that the cross term $\phi_+(y)\phi_-(y)$ can be neglected vs the diagonal terms $\phi_\pm^2(y)$. To further determine $\mathcal{N}$ for high $p$, we note that the maximum value of $e^{-a(s)p}$ is located at $s=\epsilon$ and expand $a(s)$ about that point,
\begin{equation}\label{eq:as_series}
a\left(s\right)=a(\epsilon)+\frac{\epsilon}{a(\epsilon)}(\epsilon-s)+\cdots.
\end{equation}
Therefore,
\bal
\label{eq:2.22}
\mathcal{N}\to& 2 e^{-2a(\epsilon)p}\int_0^\epsilon \mathrm{d}s\, e^{-\frac{2\epsilon p}{a(\epsilon)}(\epsilon-s)}\nn
\to & \frac{a(\epsilon)}{\epsilon p}e^{-2a(\epsilon)p}.
\eal
This yields Eq.~\eqref{eq:high-p}.

To prove that the high-pressure solution of Eq.~\eqref{eq:eigenvalue_equation} is given by Eq.~\eqref{eq:high-p}, we note that
\bal
\label{eq:integral_I}
J_\pm(y_1) \equiv &\int\mathrm{d}y_2\, e^{-a(y_1-y_2)p}\phi_\pm (y_2)  \nonumber \\
=&\int_{0}^{\epsilon}\mathrm{d}s\, \phi_\pm(y_1\mp s)e^{-a(s)p}\nonumber \\
\to&e^{-a(\epsilon)p}\int_{0}^{\epsilon}\mathrm{d}s\, \phi_\pm(y_1\mp s)e^{-\frac{\epsilon p}{a(\epsilon)}(\epsilon-s)}.
\eal
In the first step, the change in the variable $s=\frac{\epsilon}{2}\pm y_2$ has been performed, while Eq.~\eqref{eq:as_series} has been used in the second step.
Next, we expand the function $a(y_1\mp s\pm \frac{\epsilon}{2})$ appearing in $\phi_\pm(y_1\mp s)$ about $s=\epsilon$, i.e.,
\beq
a(y_1\mp s\pm \frac{\epsilon}{2})=a\left(y_1\mp\frac{\epsilon}{2}\right)\mp\frac{y_1\mp\frac{\epsilon}{2}}{a\left(y_1\mp\frac{\epsilon}{2}\right)}(\epsilon-s)+\cdots,
\eeq
so that
\bal
\label{eq:2.25}
\phi_\pm(y_1\mp s)\to & \phi_\mp(y_1)e^{\pm\frac{y_1 \mp \frac{\epsilon}{2}}{a\left(y_1 \mp \frac{\epsilon}{2}\right)}(\epsilon-s)p}\nn
\to&  \phi_\mp(y_1)e^{-\frac{ \epsilon p}{a(\epsilon)}(\epsilon-s)}.
\eal
In the second step, we have located  the function accompanying $\phi_\mp(y_1)$ at $y_1=\mp\frac{\epsilon}{2}$.
Inserting Eq.~\eqref{eq:2.25} into Eq.~\eqref{eq:integral_I} and integrating, we finally arrive at
\bal
\label{eq:}
J_\pm(y_1)\to& \phi_\mp(y_1)e^{-a(\epsilon)p}\int_{0}^{\epsilon}\mathrm{d}s\, e^{-\frac{2\epsilon p}{a(\epsilon)}(\epsilon-s)}\nonumber \\
\to & \phi_\mp(y_1)e^{-a(\epsilon)p}\frac{a(\epsilon)}{2\epsilon p}.
\eal
Therefore, in the limit $p\to\infty$, $J_\pm(y_1)\propto \phi_\mp(y_1)$. This proves that Eq.~\eqref{eq:phi_high} satisfies Eq.~\eqref{eq:eigenvalue_equation} in that limit, with $\ell$ given by Eq.~\eqref{eq:l_high}.

As a consistency test, let us reobtain Eq.~\eqref{eq:l_high} from Eq.~\eqref{eq:2.5},
\beq
\label{eq:2.28}
\ell\to \frac{2}{\mathcal{N}}\int\mathrm{d}y_1\int\mathrm{d}y_2\, e^{-a(y_1-y_2)p}\phi_+(y_1)\phi_-(y_2),
\eeq
where we have taken into account that, in the limit $p\to\infty$, the integrand is highly maximized when $y_1$ is close to $\frac{\epsilon}{2}$ and $y_2$ is close to $- \frac{\epsilon}{2}$, or vice versa.
By expanding $a(y_1-y_2)$, $a(y_1+\frac{\epsilon}{2})$, and $a(y_2-\frac{\epsilon}{2})$ around $y_1-y_2=\epsilon$, $y_1=\frac{\epsilon}{2}$, and $y_2=-\frac{\epsilon}{2}$, respectively, one has
\bal
a(y_1-y_2)+a\left(y_1+\frac{\epsilon}{2}\right)+a\left(y_2-\frac{\epsilon}{2}\right)\nn
\to 3a(\epsilon)+\frac{2\epsilon}{a(\epsilon)}(\epsilon-y_1+y_2)+\cdots.
\eal
Therefore,
\bal
\label{eq:2.30}
\ell\to & \frac{2}{\mathcal{N}}e^{-3a(\epsilon)p}\int\mathrm{d}y_1\int\mathrm{d}y_2\, e^{-\frac{2\epsilon p}{a(\epsilon)}(\epsilon-y_1+y_2)}\nn
=&\frac{2}{\mathcal{N}}e^{-3a(\epsilon)p}\left[\int\mathrm{d}y\,e^{-\frac{2\epsilon p}{a(\epsilon)}\left(\frac{\epsilon}{2}-y\right)}\right]^2
\nn
\to &\frac{2}{\mathcal{N}}e^{-3a(\epsilon)p}\left[\frac{a(\epsilon)}{2\epsilon p}\right]^2.
\eal
Taking into account Eq.~\eqref{eq:2.22}, the result \eqref{eq:l_high} is recovered.

Let us now look into the high-pressure equation of state. By using the same steps as in Eqs.~\eqref{eq:2.28} and \eqref{eq:2.30}, the integral defined by Eq.~\eqref{eq:I} becomes
\bal
\label{eq:2.31}
I\to & \frac{2}{\mathcal{N}}e^{-3a(\epsilon)p}\int\mathrm{d}y_1\int\mathrm{d}y_2\, e^{-\frac{2\epsilon p}{a(\epsilon)}(\epsilon-y_1+y_2)}\nn
&\times \left[a(\epsilon)+\frac{\epsilon}{a(\epsilon)}(\epsilon-y_1+y_2)\right]\nn
\to& a(\epsilon)\ell+\frac{2}{\mathcal{N}}e^{-3a(\epsilon)p}\frac{2\epsilon}{a(\epsilon)}\left[\int\mathrm{d}y\,e^{-\frac{2\epsilon p}{a(\epsilon)}\left(\frac{\epsilon}{2}-y\right)}\right]\nn
&\times\left[\int\mathrm{d}y\,\left(\frac{\epsilon}{2}-y\right)e^{-\frac{2\epsilon p}{a(\epsilon)}\left(\frac{\epsilon}{2}-y\right)}\right]\nn
\to &a(\epsilon)\ell+\frac{2}{\mathcal{N}}\frac{e^{-3a(\epsilon)p}}{p}\left[\frac{a(\epsilon)}{2\epsilon p}\right]^2\nn
\to& \ell\left[a(\epsilon)+\frac{1}{p}\right].
\eal
Insertion into Eq.~\eqref{eq:z_exact_01} yields Eq.~\eqref{eq:Z_high}.

\section{Numerical details}\label{app:numerical_methods}
\begin{figure}
     \centering
         \includegraphics[width=0.8\columnwidth]{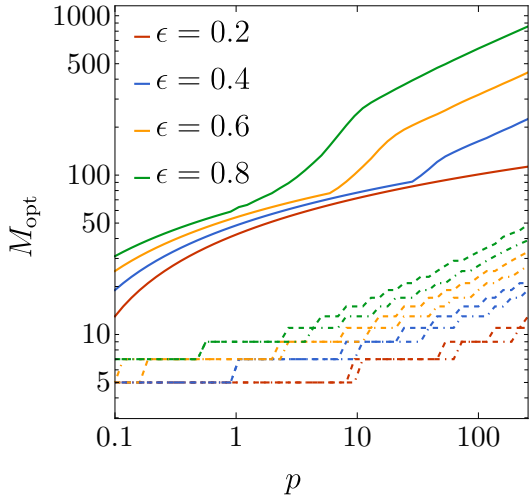}
         \caption{Optimal value ($M_{\text{opt}}$) of the number of discretization points for the exact solution (solid lines), the UPA (dashed-dotted lines), and the EPA (dashed lines) as a function of $p$ for different values of $\epsilon$.}
         \label{fig:convergence_M}
\end{figure}

In order to solve  Eq.~\eqref{eq:eigenvalue_equation} numerically, we discretize $\phi(y)$ into $M-1$ intervals, each one of size $\delta y=\epsilon/(M-1)$ [see Eq.~\eqref{eq:delta_y}],  which implies $\phi_i\equiv \phi(y_i)$, $i=1,2,\ldots,M$.
Therefore, Eq.~\eqref{eq:eigenvalue_equation} becomes
\begin{equation}
    \sum_{j=1}^{M} K_{ij}\phi_j = \ell \phi_i, \quad K_{ij} \equiv  \delta y\, e^{-a_{ij}p},
\end{equation}
or, equivalently,
\begin{equation}\label{eq:num_eigensystem}
    \mathsf{K}\cdot \bm{\phi} = \ell \bm{\phi},
\end{equation}
where $\mathsf{K}$ is  the $M\times M$ matrix of the $K_{ij}$, which is symmetric,  and $\bm{\phi}$ is the vector of  $\phi_i$. The solution of Eq.~\eqref{eq:num_eigensystem} was obtained by using standard eigensolver routines for self-adjoint matrices from the C++ EIGEN library, and then extracting the largest eigenvalue $\ell$ and its corresponding (normalized) eigenvector $\bm{\phi}$. Once obtained the solution, the compressibility factor is computed as
\begin{equation}\label{eq:num_ex}
      Z= 1 + \frac{p}{\ell} (\delta y)^2\sum_{i=1}^{M}\sum_{j=1}^{M} \, e^{-a_{ij}p}a_{ij}\phi_i\phi_j.
\end{equation}
An open-source C++ code to solve Eq.~\eqref{eq:num_eigensystem} and evaluate Eq.~\eqref{eq:num_ex} can be accessed from Ref.~\onlinecite{SingleFile}.

In the case of our approximations [see Eqs.~\eqref{Z_UPA} and \eqref{eq:z_EPA}], there is no need to solve Eq.~\eqref{eq:num_eigensystem}. The corresponding compressibility factor may be computed as
\begin{subequations}
\begin{equation}
    Z_{\mathrm{UPA}} = 1+p\frac{\sum_{i=1}^{M} \, a(s_i)(\epsilon-s_i)e^{-a(s_i)p}}{\sum_{i=1}^{M}  (\epsilon-s_i)e^{-a(s_i)p}},\quad s_i\equiv (i-1)\delta y,
\end{equation}
\begin{equation}\label{eq:num_EPA}
      Z_{\text{EPA}} = 1 + p \frac{\sum_{i=1}^{M}\sum_{j=1}^{M} \, e^{-a_{ij}p}a_{ij}\phi_{+,i}\left(\phi_{+,j}+\phi_{-,j}\right)}{\sum_{i=1}^{M}\sum_{j=1}^{M} \, e^{-a_{ij}p}\phi_{+,i}\left(\phi_{+,j}+\phi_{-,j}\right)},
\end{equation}
\end{subequations}
where $\phi_{\pm,i}\equiv\phi_{\pm}(y_i)$. However, we used, instead, the Gauss--Kronrod quadrature formula,\cite{GaussKronrod}
as implemented in the C++ BOOST library.

In the transfer-matrix solution and in our two approximations, we chose $M=\text{odd}$, so that the middle point $y_i=0$ with $i=(M+1)/2$ was included. In the three cases, the optimal value $M=M_{\text{opt}}$ was selected by the condition that the relative difference between $Z^{(M_{\text{opt}})}$ and $Z^{(M_{\text{opt}}-2)}$ was smaller than $10^{-6}$, where $Z^{(M)}$ denotes the compressibility factor evaluated with $M$ discretization points.
This optimal value is plotted in Fig.~\ref{fig:convergence_M} as a function of $p$ for some representative values of $\epsilon$. It can be seen that  $M_{\text{opt}}$ increases in the three cases with the increasing pressure and increasing pore width. Regardless of this, it is    quite apparent that $M_{\text{opt}}$ is typically an order of magnitude smaller in the UPA and EPA than that in the transfer-matrix solution. We have observed that the disparity in the values of  $M_{\text{opt}}$ becomes more pronounced as the tolerance in the relative error decreases.


    \bibliography{C:/AA_D/Dropbox/Mis_Dropcumentos/bib_files/liquid}


\end{document}